\documentclass[12pt]{article}

\usepackage{setspace}
\usepackage{amsmath}
\usepackage{psfrag,epsf}
\usepackage{graphicx} 
\usepackage{bmpsize}
\usepackage{enumerate}
\usepackage{natbib}
\usepackage{url} 
\usepackage{lineno} 
\usepackage{amssymb}
\usepackage{lineno}

\newcommand*{\affaddr}[1]{#1} 
\newcommand*{\affmark}[1][*]{\textsuperscript{#1}}

\usepackage[verbose,letterpaper,tmargin=2.54cm,bmargin=2.54cm,lmargin=2.54cm,rmargin=2.54cm]{geometry}

\begin{document}
\begin{spacing}{1.9}

\date{}
{\Large\bfseries Flexible discrete space models of animal movement}
\begin{center}
{\large Ephraim M. Hanks\affmark[1,2 *] and David A. Hughes\affmark[2,3]}\


\affaddr{\affmark[1]  Department of Statistics, The Pennsylvania State University }\\
\affaddr{\affmark[2]  Center for Infectious Disease Dynamics, The Pennsylvania State University }\\
\affaddr{\affmark[3]  Department of Entomology, The Pennsylvania State University}\\
\affaddr{\affmark[*]  hanks@psu.edu}\\

\end{center}

{\large \textbf{Summary}}
\begin{enumerate}
\item Movement drives the spread of infectious disease, gene flow, and other critical ecological processes. To study these processes we need models for movement that capture complex behavior that changes over time and space in response to biotic and abiotic factors.  Penalized likelihood approaches, such as penalized semiparametric spline expansions and LASSO regression, allow inference on complex models without overfitting. 
\item Continuous-time Markov chains (CTMCs) have been recently introduced as a flexible discrete-space model for animal movement.  Modeling with CTMCs involves discretizing an animal's path to the resolution of a raster grid.  The resulting stochastic process model can easily incorporate environmental and other covariates, represented as raster layers, that affect directional bias and overall movement rate.  
\item We introduce a weighted likelihood approach that allows for modeling movement using CTMCs, with path uncertainty due to missing data modeled by imputing continuous-time paths between telemetry locations.  The framework we introduce allows for inference on CTMC movement models using existing software for fitting Poisson regression models, including penalized versions of Poisson regression.
\item The result is a flexible, powerful, and accessible framework for modeling a wide range of animal movement behavior.
\end{enumerate}

\noindent%

\vspace{1in}
\emph{Acknowledgements:} Support for this work was provided by NSF EEID 1414296.  The authors thank Lauren Quevillon, Jeremy Sterling, Devin Johnson, and others in the Hughes Lab and the NOAA National Marine Mammal Lab for providing movement data.  
\vfill


\section{Introduction}

Animal movement is a fundamental process underlying the spread of
information, genetic material \citep{McRae2006},
infectious disease \citep{KeelingRohani}, and invasive species.
Movement is a complex combination of behaviors, and often exhibits
changing behavior over time and in response to biotic and abiotic
drivers.  Movement is typically observed as telemetry data: discrete
points in space and time marking an animal's location, often with
observational error.  In contrast, environmental factors that could
influence movement behavior often are available in gridded form, with one value
for each environmental factor at each grid cell (i.e. forest cover or elevational gradient).  

A promising approach to
pairing point-referenced telemetry data with gridded covariates is to
model movement only at the discretized resolution at which gridded
environmental factors are observed \citep{Hooten2010, Hanks2015aoas,
  Avgar2016}.  \cite{Hooten2010} modeled movement as a dynamic
occupancy process where an animal moved through a succession of
bordering grid cells.  \cite{Avgar2016} use step selection functions
to model both resource selection (preference for some combinations of landscape
characteristics) and directional persistence (correlated random walk behavior).
\cite{Hanks2015aoas} propose a stochastic process model for movement
based on a continuous-time Markov chain (CTMC).  

Both the approaches of \cite{Hanks2015aoas} and
\cite{Avgar2016} are appealing in that movement parameters describing
directional persistance and resource
selection can be fit
within a generalized linear model (GLM) framework.  This leads to
extremely efficient computing, and allows complex movement behavior to
be modeled with relative ease.  The CTMC approach of
\cite{Hanks2015aoas} is especially appealing because of the ease with
which a wide variety of movement behavior can easily be modeled,
including directional bias in movement, changes in overall movement
rate, and directional persistance.  

As movement models become more and more complex, model comparison and selection are critical to avoid overfitting the data.  Penalized likelihood methods \citep[e.g.,][]{HootenHobbs2015} provide a framework in which highly complex, sometimes semiparametric, relationships between predictors (i.e. environmental drivers) and response (i.e. observed movement) can be estimated.  Penalized likelihood approaches include information criteria based model selection, ridge regression, LASSO regression, penalized spline fitting, generalized additive models (GAMs), and more.  One major hurdle for using penalized likelihood approaches in movement analyses is the prevalence of missing data, which can occur when observation equipment fails, or (more generally) when there are gaps in observations where the animal's movement path between observations is not clear.  Within a Bayesian context, missing movement data can be accounted for by integrating over the uncertainty in the movement path between observations \citep{Johnson2008,Hooten2010,Hanks2011plos,Hanks2015aoas}.  However, this typically requires computationally intensive Markov chain Monte Carlo (MCMC) methods.  The most common likelihood-based approach to account for missing data is multiple imputation \citep[MI: ][]{Rubin1987,Rubin1996}, in which multiple stochastic realizations of the missing data are imputed from a carefully specified imputation distribution, the desired statistical model is fit to each imputed set of data in turn, and results from these multiple model fits are averaged post-hoc.  While there have been some attempts to apply penalized likelihood approaches to multiple imputation \citep{Chen2013}, these approaches only allow for point estimates, without a full characterization of uncertainty. 

Our goal in this manuscript is to (1) present discrete space CTMC models as a flexible
 framework for modeling animal movement, (2) introduce a novel stacked weighted likelihood (SWL) approach for inference in the presence of missing data, and (3) show that combining these leads to a powerful and accessible approach to modeling complex movement behavior using existing GLM software.  In
Section 2 we introduce the CTMC as a stochastic process, show how \cite{Hanks2015aoas}
used CTMCs to model movement, and describe how a wide variety of
common animal movement behaviors can be modeled within a CTMC
framework.  In Section 3 we review existing approaches for statistical
inference on movement parameters in a CTMC, and introduce our SWL approach for inference in the presence of missing
observations.  In Section 4, we provide two data examples that 
illustrate how this provides a flexible framework for understanding
and simulating complex movement behavior.  In the first, we model changing behavior
over time in a northern fur seal (\emph{Callorhinus ursinus}).  In the second, we model
 spatially-varying within nest movement behavior that varies among
 classes of the common black carpenter ant (\emph{Camponatus pennsylvanicus}).  We conclude in Section 5 with a discussion.


\section{CTMC Models for Movement on a Landscape Grid}

A continuous time Markov chain (CTMC) is a stochastic process $X(t)$ defined in continuous time and
on a discrete space.  We follow \cite{Hanks2015aoas} in our description of this stochastic process.  A complete treatment can be found in multiple textbooks \citep[e.g.,][]{RossText,KulkarniText}.  In the context of animal movement, let $X(t)$ be
the location of an animal at time $t$, where $X(t)$ can only take on one
  of $N$ distinct values: $X(t) \in \{1,2,\ldots,N\}$.  We will assume for simplicity that the $N$ locations are the $N$ grid cells on a raster representation of the animal's spatial domain, though any graphical structure is possible.  

An animal's continuous time movement path can be defined by the sequence of locations  $\{c_1,c_2,\ldots,c_T\}$ that the animal passes through, which is called the \emph{embedded chain}, and the \emph{transition times} at which the animal moves between locations $\{t_1,t_2,\ldots,t_T\}$.  The transition times can also be equivalently represented by the \emph{residence times} $\{\tau_1,\tau_2,\ldots,\tau_T\}$, where $\tau_k=t_{k+1}-t_k$ is the time spent in $c_k$, the $k$-th location visited by the animal.   

A CTMC statistical model for such a path is written in terms of the transition rates $\{\alpha_{ij}\geq 0,\  i\neq j;\  i=1,\ldots,N;\
  j=1,\ldots,N\}$ which are parameters that control movement behavior between
  spatial locations (grid cells).  If it is impossible to directly move from location $i$
  to location $j$, then $\alpha_{ij}=0$.  Higher rates between
  locations correspond to increased rate of movement between those locations. 
%
If the $k$-th location in the embedded chain is $i$ ($c_k=i$), and the animal's movement follows a CTMC with transition rates $\{\alpha_{ij}\}$, then the time $\tau_k$ the animal will remain in location $i$ is
  exponentially-distributed with rate equal to sum of all $\alpha_{ij}$
  with first index $i$ - all rates ``leaving'' location $c_k=i$:
\begin{equation}
\tau_k=(\text{residence time in node }c_k=i) \sim \text{Exp}\left(\sum_{j=1}^{N} \alpha_{ij}\right).
\end{equation} 
So the mean residence time in node $i$ is $E(\tau_k|c_k=i)=1/\sum_{j=1}^{N} \alpha_{ij}$.

When an animal leaves node $c_k=i$ and transitions to some neighboring
  spatial location, the probability that the transition is to node $c_{k+1}=\ell$ is
\begin{equation}
P(c_{k+1}=\ell | c_k=i)=\frac{\alpha_{i\ell}}{\sum_{j=1}^{N} \alpha_{ij}}.
\end{equation}


Consider the case where we have two sets of covariates, one set which we assume is related to an animal's absolute speed and another set which we assume is related to directional bias.  We can use these two sets of covariates to model the transition rates of a CTMC, as follows.  Let ($\textbf{m}_1,\textbf{m}_2,\ldots,\textbf{m}_L$) be $L$ raster covariates, where $m_{\ell k}$ is the value of the $\ell$-th covariate at the $k$-th grid cell, and $\mathbf{m}_\ell$ denotes a column vector of these values at each of the $N$ grid cells.  We assume that these $L$ covariates affect absolute speed (motility), and will call these covariates ``motility covariates''.  Additionally, assume that we
have $P$ raster covariates
($\mathbf{d}_1,\mathbf{d}_2,\ldots,\mathbf{d}_P$) that we assume
affect the directional bias of animal movement, through the gradient of a potential surface \citep{Brillinger2001,Hanks2011plos,Preisler2013}.  
The gradient of
$\mathbf{d}_p$ at ($x_m,y_m$) is a vector pointing in the direction of
steepest increase in $\mathbf{d}_p$ at that point in space.  To model
directional bias, first define a vector $\mathbf{e}_{ij} =
(x_j-x_i,y_j-y_i)$ for each transition rate
$\alpha_{ij}$.  Under this formulation, $\mathbf{e}_{ij}$ points in
the direction that an animal moves if it transitions from grid cell
$i$ to grid cell $j$ in the CTMC model.  The dot product 
\begin{equation}
q_{pij}=\mathbf{e}_{ij}'\mathbf{g}_{pi}=(x_j-x_i)g_{pmx}+(y_j-y_i)g_{pmy}
\end{equation}
 is positive if
$\mathbf{e}_{ij}$ (the direction of movement from node $i$ to $j$) and
$\mathbf{g}_{pi}$ (the direction of steepest ascent of the covariate
$\mathbf{d}_p$ at node $i$) point in the same direction, zero if
$\mathbf{e}_{ij}$ and $\mathbf{g}_{pi}$ are at right angles to each
other, and negative if they point in opposite directions.  The dot product $q_{pij}=\mathbf{e}_{ij}'\mathbf{g}_{pi}$ provides a directional covariate that captures the correspondence between a potential movement and the gradient of a covariate, and can be used to model movement bias in the direction of increasing (or decreasing) levels of a covariate.

The CTMC is an appealing framework for jointly modeling variation in absolute movement rate and directional bias, as both motility and directional covariates can be integrated into a loglinear model for the CTMC transition rates $\alpha_{ij}$
\begin{equation}
\alpha_{ij}=\begin{cases}
\exp \left\{\sum_{k=1}^L \delta_k m_{ki} + \sum_{p=1}^{P}
\gamma_p\mathbf{e}_{ij}'\mathbf{g}_{pi}\right\}  &,\text{ if }i\text{ and }j\text{ are neighbors}\\
0 &,\text{ otherwise}
\end{cases}.  
\end{equation}
 
Consider the simplified case in which $\gamma_p=0,p=1,2,\ldots,P$ and the transition rates are $\alpha_{ij}=\exp\{\sum_k\gamma_k\ell_{ki}\}$.  The overall rate of movement out of cell $i$ is $\sum_j \alpha_{ij}$, and so is defined by the motility covariates ($\textbf{m}_1,\textbf{m}_1,\textbf{m}_L$) and their coefficients $\boldsymbol\delta$.  If grid cells $u$ and $v$ are both neighbors of $i$, then $\alpha_{iu}=\alpha_{iv}$, so there is no directional bias in movement.  From this we see that the motility covariates ($\textbf{m}_1,\textbf{m}_1,\textbf{m}_L$) only affect absolute rate of movement.

When the $\gamma_p$ are not all zero, the transition probabilities (2), which define directional bias in movement, are independent of the motility covariates ($\textbf{m}_1,\textbf{m}_1,\textbf{m}_L$), as 
\[\alpha_{ij}=(\exp\{\sum_k \delta_k \ell_{ki}\})(\exp\{\sum_{p}
\gamma_p\mathbf{e}_{ij}'\mathbf{g}_{pi}\})\]
 and the transition probabilities (2) are
\begin{equation}
P(c_{k+1}=u | c_k=i)=\frac{\alpha_{iu}}{\sum_{j=1}^{M} \alpha_{ij}}=\frac{\exp\{\sum_{p}
\gamma_p\mathbf{e}_{iu}'\mathbf{g}_{pi}\}}{\sum_j \exp\{\sum_{p}
\gamma_p\mathbf{e}_{ij}'\mathbf{g}_{pi}\}}.
\end{equation}
The result is that a CTMC model on a raster grid with rates defined by (4) provides a flexible general framework for jointly modeling covariate effects 
on absolute movement rate and directional bias.  

\subsection{Modeling Movement Using Covariate Rasters}

A wide variety of movement behavior
can be modeled by careful specification of covariate raster layers.  Examples include modeling migration along environmental gradients \citep{Hooten2010}, central-place foraging behavior \citep{Hanks2011plos,Hanks2015aoas}, and interaction with conspecifics \citep{Hanks2015aoas,Quevillon2015}.  We now provide examples of modeling common movement behaviors using raster layers to define CTMC movement rates (4).

\textbf{\emph{Movement along environmental gradients.}}  
Animals may use environmental gradients for navigation.  For
example,
a mule deer might move 
predominantly in the direction of increasing elevation during a
spring
migration \citep[e.g.,][]{Hooten2010}, or a seal might follow
gradients in sea surface temperature
to navigate toward land \citep[e.g.,][]{Hanks2011plos}.  This could be modeled by using a raster layer $\mathbf{d}_p$ of the relevant environmental variable and then using the gradient (3) in the loglinear model for CTMC transition rates (4).

\textbf{\emph{Activity centers.}} The gradient of Euclidean distance to the animal's ``central place'' (e.g., den or rookery) could be used as a gradient-based covariate (3).  A negative regression coefficient $\gamma_p$ in (4) would then lead to movement biased to return to the central place.

\textbf{\emph{Environmentally driven movement rate.}}  Many animals tend to move more quickly through unfavorable terrain (e.g., when crossing a road).  In other situations, animals may move more cautiously and slowly through unfavorable terrain (e.g., when entering an open field).  An indicator raster layer for each relevant cover type could be used as a motility covariate $\textbf{m}_k$ in the model for CTMC transtion rates (4), with positive values of the corresponding regression coefficient $\delta_k$ indicating an increase in overall movement rate through the $k$-th land cover type. 

\textbf{\emph{Conspecific Interaction.}}  An animal's movement patterns can be greatly affected by interaction
with conspecifics.  When inference is desired on the movement path of one focal animal, and there are observed paths of other nearby animals, a gradient-based covariate (3) could be created from the time-varying distance from each conspecific to the focal animal, thus modeling attraction to (with $\gamma_p<0$) or repulsion from (with $\gamma_p>0$) the conspecific.

\textbf{\emph{Directional persistence.}}  To model correlated random walk behavior within the Markovian CTMC framework, \cite{Hanks2015aoas} include an autocovariate in the loglinear model for $\alpha_{ij}$, where the autocovariate is created from the gradient covariate as in (3) where the gradient vector $\mathbf{g}_{pi}$ is a vector of the animal's current direction of movement from the most recent grid cell in the embedded chain to the current grid cell.  Positive values of the regression coefficient related to this autocovariate indicate that the
animal is likely to maintain its direction of movement over time.

\textbf{\emph{Memory.}}  Including a raster covariate of distance to past locations as a gradient-based covariate (3) models memory through attraction to spatial locations visited in the past.

\textbf{\emph{Individual heterogeneity.}}  Including a random intercept in the loglinear model (4) would model individual heterogeneity in overall movement rates.  Similarly, an individual-specific random slope model for a regression coefficient in (4) would model individual variation in response to that raster covariate.

\textbf{\emph{Changing behavior over time.}}  \cite{Hanks2015aoas} use varying-coefficient models \citep{Hastie1993} to model periodic (i.e., seasonal or diurnal) variation in movement response to covariates.  In a varying-coefficient model, a static in time term $\delta_k\ell_{ki}$ in (4) is replaced by $\delta_k(t)\ell_{ki}$, where $\delta_k(t)$ is a time-varying coefficient often modeled smoothly using a penalized spline basis expansion.

\section{Statistical Inference for CTMC Model Parameters}



The likelihood of the observed CTMC path
($\mathbf{c},\boldsymbol\tau$), defined by the \emph{embedded chain}
$\mathbf{c}$, and the \emph{residence times} $\boldsymbol\tau$, is given by the product of the density of
the embedded chain (2) and the residence times (3)
 \begin{align}
f(\mathbf{c},\boldsymbol\tau|\boldsymbol\delta,\boldsymbol\gamma)&=\prod_{t=1}^{T} \left[ \left(\sum_{k=1}^M
\alpha_{c_t k}\right) \exp\left\{ - \tau_t\sum_{k=1}^M \alpha_{c_t k}\right\}\right]\cdot \left[ \prod_{t=1}^{T}
\frac{\alpha_{c_tc_{t+1}}}{\sum_{k=1}^M \alpha_{c_t k}} \right] 
\end{align}
with transition rates a function of raster covariates and regression parameters $\{\boldsymbol\delta,\boldsymbol\gamma\}$ as in (4).

\cite{Hanks2015aoas} introduced a set of auxilliary variables $\{z_{tk}\}$ to facilitate maximizing (6).  For each grid cell $c_t$ in the embedded chain, create
$\{z_{tk},k=1,\ldots,N\}$ defined as indicator variables for the embedded chain, so that each $z_{tk}$ is zero except for $z_{tc_{t+1}}$, the latent variable corresponding to $c_{t+1}$, the next grid cell in the
embedded chain 
\begin{equation}
z_{tk}=\begin{cases}
1 &,\text{ if } k=c_{t+1}\\
0 &,\text{ if } k\neq c_{t+1}
\end{cases}.
\end{equation}
\cite{Hanks2015aoas} showed that maximizing the likelihood (6) of the CTMC path is identical to maximizing the 
likelihood of a Poisson
regression with canonical log link, where $z_{tk}$ as the response,
$\mathbf{x}_{c_tk}$ are the covariates in the linear predictor, and
$\log(\tau_t)$ is an offset
\begin{equation}
z_{tk}\stackrel{iid}{\sim} \text{Pois}(\tau_t \alpha_{c_tk}), \ \ t=1,2,\ldots,T;\ \ k=1,2,\ldots,N
\end{equation}
with $\{\alpha_{c_t k}\}$ as in (4).  The likelihood of (8) is given by
\[f(\mathbf{z},\boldsymbol\tau|\boldsymbol\beta) \propto \prod_{t=1}^{T} \prod_{k=1}^{M} \left(\alpha_{c_t k}^{z_{tk}}
   \exp\left\{ - \tau_t \alpha_{c_t k} \right\} \right) \]
which, with some algebraic manipulation, is equivalent to (6).  This means that the CTMC parameters in (4) can be estimated using standard Poisson GLM software by first creating the $\{z_{tk}\}$ from the embedded chain $\mathbf{c}$ and then fitting the model (8) using existing software.

\subsection{Inference from Discrete Observation in Time}

Instead of observing a complete CTMC path $(\mathbf{c},\boldsymbol\tau)$, it is typical to observe an animal's location at $K$ discrete points in time, leading to data in the form of
$\{(x_i,y_i,t_i),\ t_i=t_1,t_2,\ldots,t_K\}$.  Once discretized to
raster cells $\{c_i, \ i=1,2,\ldots,K\}$, the likelihood of the observed data is 
\[\prod_{i=1}^K \mathbb{P}_{c_k c_{k+1}}(t_{k+1}-t_k)\]
where $\mathbb{P}_{ij}(\Delta t) = P(X(t+\Delta t)=j | X(t)=i)$.
Obtaining these transition probabilities for a fixed time $\Delta t$
requires matrix exponentiation \citep{RossText} of an $N \times N$
generator matrix defined by the set of all transition rates between
nighboring grid cells.  For large raster grids this is computationally
prohibitive.  

Instead, multiple authors have considered inference based on
stochastic imputation of continuous-time continuous-space paths
linking the observed locations \citep[e.g.,][]{Johnson2008,Hooten2010,Hanks2011plos}.  
In this approach, continuous-time paths can be imputed from the posterior distribution of a movement model fit to the telemetry data.  For example, the continuous-time correlated random walk
(CTCRW) model of Johnson et al., (2008) or the functional movement model of \cite{Buderman2016} could be fit to the original telemetry data, and multiple paths could be simulated from the fitted
model.  Inference on CTMC movement parameters, conditioned on the
telemetry data, could then be obtained by integrating over the imputed
paths \citep{Johnson2008,Hooten2010,Hanks2011plos,Hanks2015aoas}.
This integration could be carried out in a Bayesian context through
sampling a new continuous-time movement path at each iteration of a
MCMC sampler, or approximated by multiple imputation when making
inference based on maximum likelihood \citep[as done
in][]{Hanks2015aoas}.  The Bayesian approach is computationally intensive, and existing penalized likelihood approaches using multiple imputation \citep{Chen2013} only allow for point estimates without quantification of uncertainty (e.g., through confidence intervals).

\subsection{A Stacked Weighted Likelihood Approach to Inference With Missing Data}

To facilitate fitting semiparametric and other complex models to movement data, we present a novel stacked weighted likelihood (SWL) approach to inference in the presence of missing data.  The SWL approach is to first impute $P$ continuous time movement paths $\{(\mathbf{z}^{(i)},\boldsymbol\tau^{(i)}),\ i=1,\ldots,P\}$ from an imputation distribution, such as the posterior predictive distribution of the functional movement model of \cite{Buderman2016}.  Under multiple imputation, a CTMC model would be fit to each path individually, and inference on $(\boldsymbol\delta,\boldsymbol\gamma)$ could be made by combining the results of each individual fit using the rules developed by \cite{Rubin1987}.  See \cite{Hanks2015aoas} and \cite{Nakagawa2010} for examples.

Instead, we propose inference based on maximizing the following likelihood, which 
is the geometric mean of the likelihoods of each imputed CTMC path
\begin{equation}
f(\{(x_i,y_i,t_i)\}|\boldsymbol\delta,\boldsymbol\tau)=\prod_{i=1}^P
\left[f(\mathbf{z}^{(i)},\boldsymbol\tau^{(i)}|\boldsymbol\delta,\boldsymbol\tau)\right]^{1/P}.
\end{equation}
Each term in the product is the likelihood (8) of one of the
imputed paths after the creation of the auxiliary $\mathbf{z}$ (7).  This formulation is a weighted
likelihood approach, similar to data cloning \citep{Lele2007,Lele2011}, where we assign a weight of $1/P$ to each of the
$P$ imputed paths, reflecting that the $P$ imputed paths should each be given equal weight, and taken all together should be given the ``weight'' of the 1 observed set of telemetry data.  As the individual terms in the SWL likelihood (9) are given by the likelihood of independent Poisson regression observations (8), the SWL approach amounts to ``stacking'' the $P$ imputed data sets $\{(\mathbf{z}^{(i)},\boldsymbol\tau^{(i)}),\ i=1,\ldots,P\}$, treating each as independent but downweighted observations.  We show in Appendix A that the inference using the SWL (9) is nearly identical to inference obtained using multiple imputation.  The SWL is appealing because of the ease with which it can be implemented, and the straightforward extension to estimation using penalized likelihood approaches, as the SWL likelihood takes on a closed form (9).

\subsection{A General Framework For Flexible Discrete Space Movement Modeling}

Pairing the SWL approach to inference with the CTMC model for animal movement provides a flexible and powerful framework that is easy to implement using existing GLM software.  We suggest the following general steps.
\begin{enumerate}
\item Collect telemetry data in the form of $\{(x_i,y_i,t_i),\
  t_i=t_1,t_2,\ldots,t_K\}$.  
\item Collect and create covariate rasters.  
  Each covariate defines either (or both) a motility covariate $\textbf{m}_k$ or a directional (gradient based) covariate $\mathbf{d}_p$ in (4).
\item Impute $P$ continuous paths for each set of animal telemetry data using a suitable continuous-time movement model.  
\item Discretize each continuous path in space to obtain $P$
  CTMC paths 
$\{(\mathbf{c}, \boldsymbol\tau)^{(1)}, (\mathbf{c},\boldsymbol\tau)^{(2)}, \ldots, (\mathbf{c},\boldsymbol\tau)^{(P)}\}$, and the corresponding covariates $\{m_{ki},\mathbf{e}'_{ij}\mathbf{g}_{pi}\}$ from (4).   
\item Convert each CTMC path to data $(\mathbf{z},\boldsymbol\tau)^{(1)},(\mathbf{z},\boldsymbol\tau)^{(2)},\ldots,(\mathbf{z},\boldsymbol\tau)^{(P)}$ appropriate for a Poisson GLM
  likelihood by creating the auxiliary variabless (7) for each path.  
\item Stack together the response $\mathbf{Z}=(\mathbf{z}^{(1)}{}'\ \mathbf{z}^{(2)}{}'\ \cdots 
\ \mathbf{z}^{(P)}{}')'$, offset $\boldsymbol\tau=(\boldsymbol\tau^{(1)}{}'\ \boldsymbol\tau^{(2)}{}'\ \cdots 
\ \boldsymbol\tau^{(P)}{}')'$, and covariates $\mathbf{X}= (\mathbf{X}^{(1)}{}' \ \mathbf{X}^{(2)}{}' \ \cdots \ \mathbf{X}^{(P)}{}')'$ from (4). 
\item Make inference on CTMC transition rate parameters $\boldsymbol\beta$ under the SWL likelihood (9), or a penalized version, using Poisson GLM software with each observation given a ``weight'' of $1/P$, thus averaging
  over the uncertainty in the unobserved parts of the movement path.  Alternately, Bayesian approaches or multiple imputation can be used for inference.
\end{enumerate}

\section{Examples}

We now illustrate using two example systems the ease with which complex movement behaviors can be modeled in this framework.

\subsection{Example 1: Northern Fur Seal Movement}

We first consider a northern fur seal (henceforth, seal) movement
path.  Northern fur seals are pelagic foragers found primarily in the
North Pacific Ocean.  During the Summer months, the northern fur seal population gathers at the Pribilof Islands off the coast of Alaska, USA to breed.  Figure 1a shows a 21-day movement path taken by a male seal who leaves the Pribilof Islands (shown with a red triangle), forages in the open Pacific ocean, and then returns to the Pribilof Islands.  Observations were obtained using the Argos system (Argos website https://www.argos-system.org), with telemetry fixes attempted every 4 hours.  The background image shows sea surface temperature (SST) in degrees Celsius.  The animal's movement path shows time-varying directional bias, as the seal first swims away from its rookery on the Pribilof Islands (Figure 1b), and then returns.  Further details on data collection are given in \cite{Hanks2011plos}.

\cite{Hanks2011plos} modeled changing behavior over time using a state-switching model which allowed for an unknown number of movement states through a reversible jump MCMC algorithm.  The velocity-based model of \cite{Hanks2011plos} is only able to model directional bias through gradient based covariates ($\mathbf{g}$ in (4)), and the reversible jump algorithm took multiple hours to converge.  Through the use of a CTMC model, we will model additional complexity in seal movement by modeling overall movement rate in different environmental conditions (different water temperatures), and show that inference using the SWL approach takes only seconds using standard GLM software.

We first fit the functional movement model of \cite{Buderman2016} to the seal telemetry data.  Four imputed paths from the resulting posterior predictive distribution are shown in Figure 1c.  We imputed a total of $P=20$ continuous-time paths, and used the `ctmcmove' package \citep{ctmcmove} in the R statistical computing environment to discretize each imputed path to a CTMC path (Figure 1d shows a detailed discritization of one path).  Code to replicate this analysis is given in Appendix A.

We modeled seal movement as a CTMC with sea surface temperature (SST) as a motility covariate (4) affecting overall rate of movement.  In addition, we modeled time-varying desire to swim away from, and then return to, the Pribilof Islands, through a directional covariate raster (Figure 1b) with values equal to the great circle distance to the Pribilof Islands.  We used a semiparametric varying-coefficient model to capture movement bias first in the direction of increasing distance from the seal's rookery, and then in the direction of decreasing distance.  This amounts to a time-varying potential function approach \citep{Preisler2013}.  We also modeled directional persistance through the inclusion of an autocovariate $\mathbf{e}'_{ij}\mathbf{a}_{t}$, where $\mathbf{a}_{t}$ is an autocovariate pointing in the direction of current movement, as described in Section 2.1.  Thus, the model we assume for the time-varying CTMC transition rates (4) between neighboring grid cells $i$ and $j$ is 
\begin{equation}
\alpha_{ij}(t)=\exp \{ \delta_0+\delta_1 \text{SST}_i + \gamma(t) \mathbf{e}'_{ij}\mathbf{g}_{i} + \rho \mathbf{e}'_{ij}\mathbf{a}_{t}\}
\end{equation}
where $\mathbf{g}_i$ is the gradient of the distance from grid cell $i$ to the Pribilof Islands, and SST$_i$ is the sea surface temperature of the $i$-th grid cell.  The time varying-coefficient $\gamma(t)$ was modeled using a semiparametric spline basis expansion using B-splines.  Smoothness in this additive model was imposed by penalizing the integral of the square of the second derivative $\int (\gamma''(t))^2 \text{d}t$ \citep[e.g.,][]{Fahrmeir2013}.  The penalized log-likelihood used for estimation was then given by 
\begin{equation}
PLL(\{(x_i,y_i,t_i)\}|\delta_0,\delta_1,\gamma(t), \rho)=\sum_{p=1}^P
\frac{1}{P} \log
\left[f(\mathbf{z}^{(p)},\boldsymbol\tau^{(p)}|\delta_0,\delta_1,\gamma(t),\rho)\right]^{1/P}-\lambda \int [\gamma''(t)]^2 \text{d}t
\end{equation}
with the tuning parameter $\lambda$ chosen via cross-validation.  The appeal of the SWL approach is that (11) can be fit using any software which fits varying coefficient models for Poisson regression.  We used the `mgcv' package in R \citep{mgcv} which implements a variety of penalized semiparametric models.  Estimation took only a few seconds, representing a significant increase in computational efficiency relative to the simpler directional-only model of \cite{Hanks2011plos}.  

The results of this analysis showed that the seal in general moves more slowly in higher temperature water ($\hat{\delta_1}=-.14$, $p$-value=.014), the seal exhibits directional persistance ($\hat{\rho}=0.461$, $p$-value $<10^{-10}$), and that the seal is first drawn away from its rookery in the Pribilof Islands, and then back toward the rookery (Figure 1e).  Figure 1f shows 3 simulated 21-day seal  movement paths from the fitted model.  These paths show similar behavior to the observed NSF movement in Figure 1a, with simulated animals making loops out and back to the rookery, and showing slower movement in regions of higher SST.  This behavior could not have been captured without modeling absolute movement rate based on covariates, directional bias based on covariates, and changing behavior over time.  Pairing the CTMC model for movement with penalized approaches for inference leads to a flexible and powerful framework for modeling complex movement behavior.

\subsection{Example 2: Semiparametric Modeling of Ant Movement}

We now illustrate how CTMC movement models can be applied to
population-level movement through the analysis of a colony of the
common black North American carpenter ant \emph{Camponatus
  pennsylvanicus}.  Ants, like those in the genus \emph{Camponatus},
provide an important laboratory species for studying collective
behavior, community space use, and social contact structure, as entire
colonies can be observed at once, something rarely possible with
humans or other vertebrates.  Our goal is to capture ant movement
behavior sufficiently that we can simulate realistic ant movement.
This would allow for future simulation of epidemics through the ant
colony.  For this analysis, we consider a single ant colony from the
study described in \cite{Quevillon2015}, in which a Queen and 16
workers were placed in a 4-chambered wooden nest measuring 7cm by
11cm.  Videos of ant movement were recorded using a GoPro Hero2 camera
with a modified IR filter (RageCams.com) illuminated under infrared
light.  As ants cannot detect infrared light, we expect ant behavior
in this constructed environment to be similar to ant behavior in
natural dark nest environments (tunnels in dead and living wood for \emph{Camponatus}).

\cite{Quevillon2015} found that, similar to human societies, different classes of ants have different roles in the colony, and exhibit different space use and social contact behavior.  Our goal of this analysis is to fit a flexible movement model that captures differences in behavior between classes of ant sufficiently that we can simulate realistic ant movement.  We first divided worker ants into two broad classes: ``foragers'', which are all worker ants seen to have left the nest in the observation window, and ``nest ants'', which are all worker ants who have not been observed leaving the nest.  Figure 2a-2c show the observed patterns of space use for each of these classes of ant, discretized to a 1cm square grid.  The Queen (Figure 2c) confines herself to a region of the top-right chamber.  The nest ants (Figure 2b) spend much of their time hear the Queen, while the foragers (Figure 1a) spend most of their time in other chambers of the nest.  This segregation in space can convey a measure of social immunity to disease \citep{Quevillon2015} as the foragers, which are at high risk of encountering pathogens while out of the nest, are buffered from contact with the Queen.

We take a purely semiparametric approach to modeling ant movement.
For each class of ant, we specify the CTMC movement rates (4) at each
1cm grid cell in the nest as a loglinear function of a motility
surface $m(x,y)$, which models spatially-varying average movement rate,
and the gradient of a potential surface $p(x,y)$, which models
spatially-varying directional bias
\begin{equation}
\alpha_{ij}=\exp \{ m(x_i,y_i)+ \mathbf{e}_{ij}' (p(x_j,y_j)-p(x_i,y_i)).
\end{equation}
Both the motility surface $m(x,y)$ and potential surface
$p(x,y)$ are modeled using a B-spline expansion ($m(x,y)=\sum_k
\delta_k \phi_k(x,y)$ and  $p(x,y)=\sum_k \gamma_k \phi_k(x,y)$).
This is similar to the approach proposed by \cite{RussellAntMove}.  We estimated $m(x,y)$ and $p(x,y)$ by using LASSO regression in which the SWC likelihood (9) is penalized by the sum of the absolute values of $\{\delta_k\}$ and $\{\gamma_k\}$.  This allows for joint penalization (and model selection) of spatially-varying movement rate ($m(x,y)$) and directional bias (defined by the potential function $p(x,y)$).  

We allowed each class of ants to have their own potential and motility
surfaces, and show the resulting estimated surfaces in Figure 2d-2i.
Ants tend to move in the direction of decreasing potential ($p(x,y)$),
so the estimated potential surface for the Queen (Figure 2f) indicates
that the Queen will reside mostly within a potential ``valley'' in the
top-right chamber of the nest.  The foragers' potential surface
(Figure 2d) shows a potential ``hill'' in the same location,
indicating that foragers will in general avoid entering the immediate
vicinity of the Queen.  The estimated motility surfaces for the
foragers (Figure 2g) and nest ants (Figure 2h) show increased motility
(movement rate) in the doorways between chambers, indicating that ants
are unlikely to rest in these locations.  This semiparametric model
provides a significant increase in our understanding of ant movement
and space use in the nest relative to the parametric analysis in \cite{Quevillon2015}.

We simulated ant movement from the CTMC model by simulating CTMC paths on the 1cm grid for a Queen and a set of foragers and nest ants matching the numbers of \emph{Camponatus} observed in each of the 8 days of observation.  The simulated space use (Figures 2j-2l) shows that the CTMC model captures space use within the nest of the three classes of ant modeled here.

\section{Discussion}

Pairing a CTMC model for movement with our proposed SWL approach for inference results in a flexible, powerful framework for modeling a wide variety of movement behavior.  In particular, the discrete-space nature of a CTMC allows for specifying raster layers of environmental (or other) factors as covariates driving variation in movement rate and directional bias.  We have provided two examples in which changing movement behavior in time and space is modeled using semiparametric models with parameters estimated using penalized likelihood approaches.  The SWL (9) allows easy model selection based on AIC, or other penalized approaches \citep{HootenHobbs2015}, and straightforward computing using existing software for fitting GLMs.

Discretizing space results in an approximation to a true continuous-space movement path.  A growing body of literature provides examples of increasingly complex continuous-space movement models \citep{Hooten2016,Russell2015,Scharf2015}.  When the discretization required for a CTMC model is undesirable, the ease with which a CTMC model can be fit makes it an appealing tool for exploratory analysis prior to fitting a more complex continuous-space model using MCMC (or something similar).

We have provided code to replicate the analyses in this paper in Appendix A and Appendix B.  Additionally, the `ctmcmove' R-package \citep{ctmcmove} provides code to facilitate the general framework for modeling movement described in Section 3.3 of this manuscript, as illustrated by the code in the Appendices.  

The straightforward inference possible within a CTMC framework for
modeling movement promises to make complicated movement modeling more
accessible and facilitate more realistic modeling.  Our two examples
illustrate that fitting flexible CTMC movement models using penalized
likelihood approaches leads to models for movement that capture many
realistic features of animal movement.  In both examples the CTMC
framework led to novel insights into animal behavior not initially
understood in earlier publications.  The ability to easily fit complex
models to telemetry data will make it possible to build more realistic
descriptions of movement behavior and ultimately increase our
knowledge of movement ecology and the important processes, like gene
flow and the spread of infectious disease, that are driven by
movement.

\bibliographystyle{ecology}
 \bibliography{C:/Users/Ephraim/Dropbox/Research/libz}

\begin{figure}
\hspace{-.65in}
\includegraphics[width=1.2\textwidth]{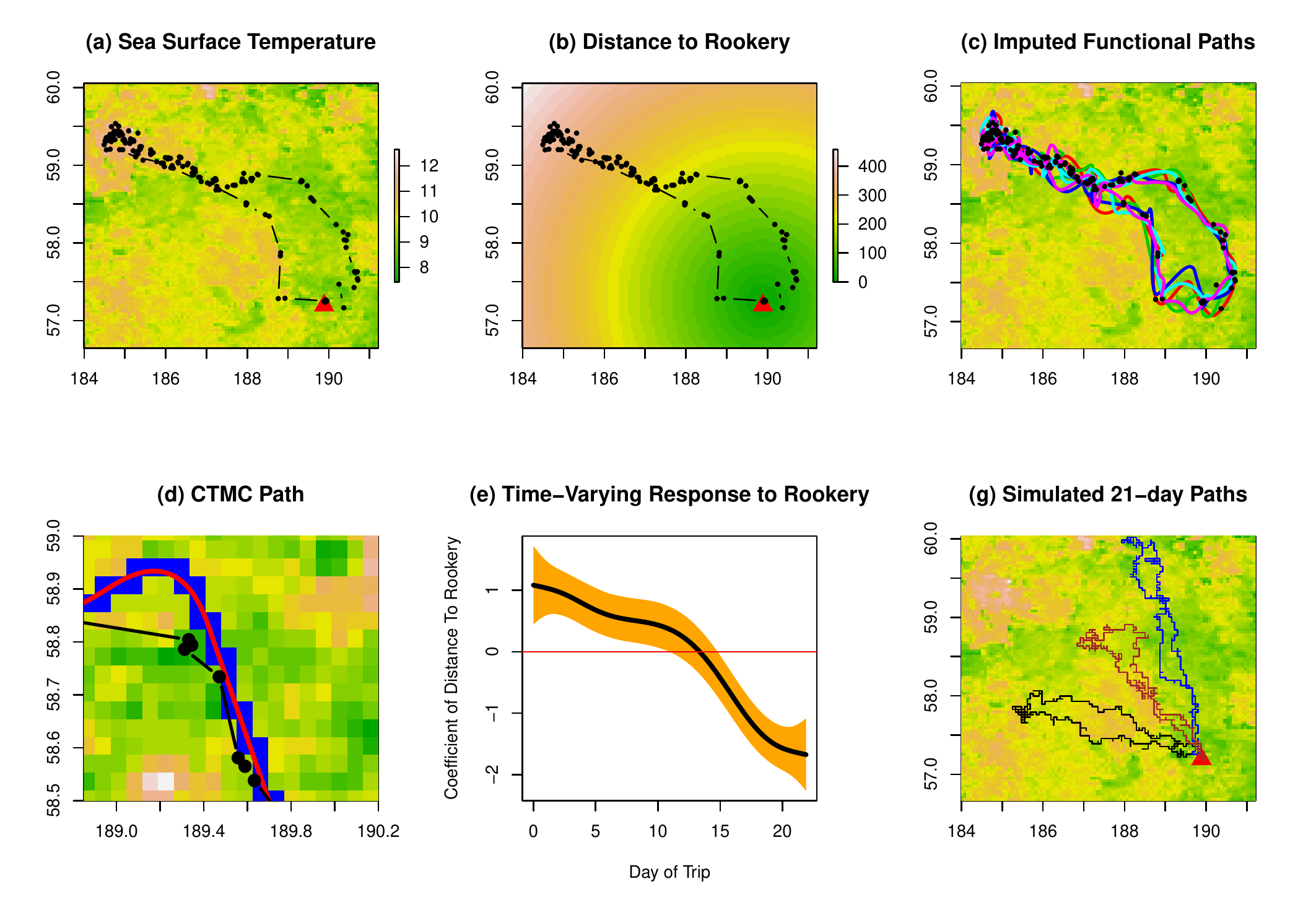}
\caption{\textbf{Discrete Space Modeling of Seal Movement}. Telemetry
  data from one 21-day foraging trip taken by a northern fur seal is
  shown on the backdrop of sea surface temperature (a) and distance
  from the animal's rookery (b) in the Pribilof islands in the North
  Pacific Ocean.  Sea surface temperature is measured in degrees
  Celsius and distanct to rookery in kilometers.  To account for uncertainty in the path between telemetry locations, we fit a continuous-time model for movement to the telemetry data.  Five realizations of paths from the posterior predictive distribution of this model path are shown in (c).  Each path is discretized to a CTMC path (d) on the raster grid.  We fit a CTMC model for seal movement behavior with a time varying-coefficient model for directional bias.  Results (e) shows the animal is first drawn away from the rookery, then returns.  Simulated paths from the fitted model (f) show that this CTMC model captures many features of the seal movement.}
\end{figure}

\begin{figure}
\hspace{-.65in}
\vspace{-.25in}
\includegraphics[width=1.2\textwidth]{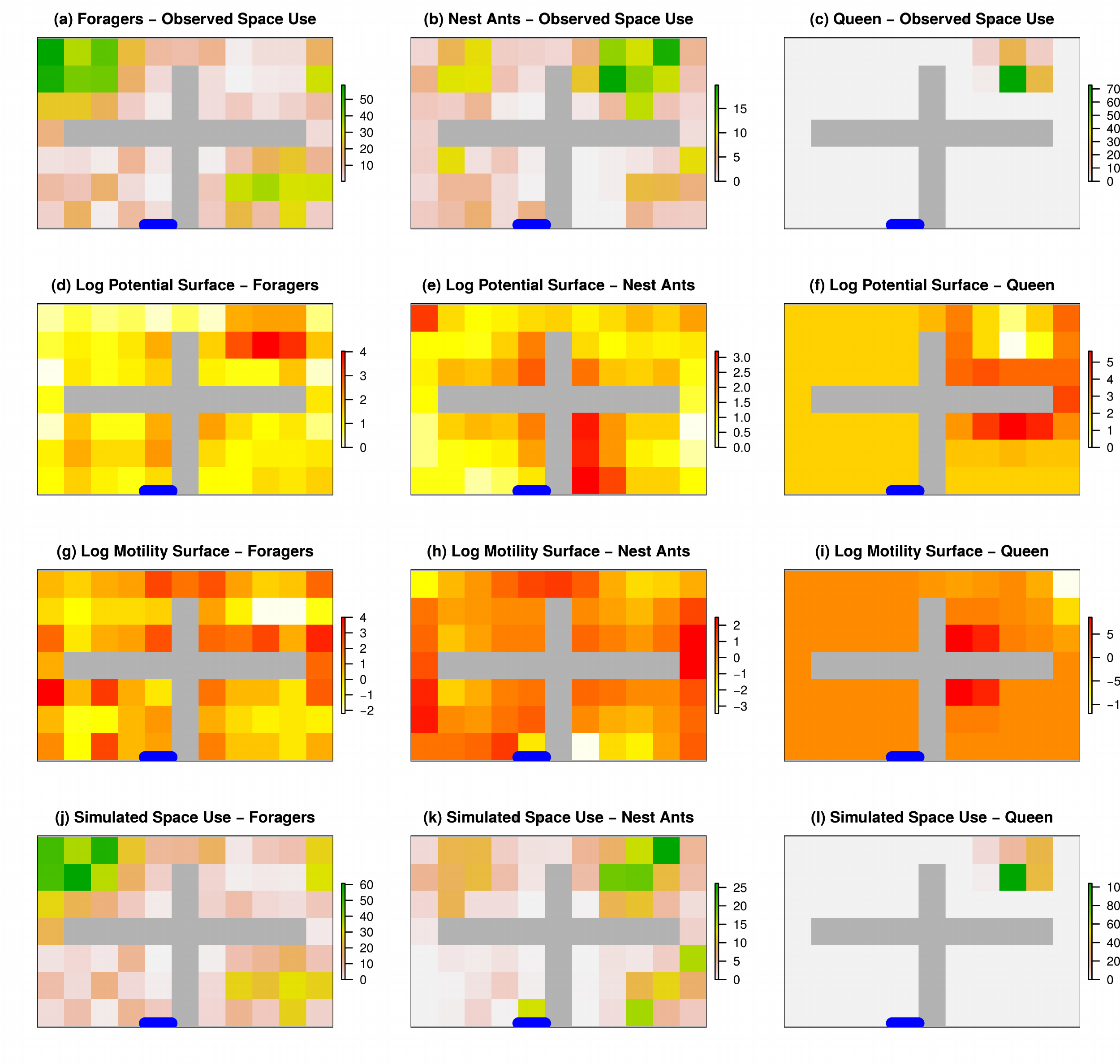}
\caption{\textbf{Semiparametric Modeling of Ant Movement}.  Movement trajectories from 17 ants in an 11cm by 7cm four-chambered nest were observed for 20 minutes for 8 days.  The entrance to the nest is shown in blue.  After discretizing these movements into 1cm grid cells, the aggregated space use (in ant-minutes) for (a) foraging ants, (b) nesting ants (inactive foragers), and (c) the Queen are shown.  After fitting a CTMC movement model with a spatially-varying semiparametric model for both motility (absolute movement rate) and movement potential (directional bias), the estimated surfaces for each classification of ant are shown in (d)-(i).  Movement trajectories simulated from the fitted CTMC for an \emph{in silico} colony show that simulated ant movement results in aggregated space use in the nest (j)-(l) that is very similar to observations (a)-(c).}
\end{figure}

\end{spacing}
\end{document}